\documentclass[jgrga]{agutex}
\usepackage{graphicx}
\authorrunninghead{TATJANA ZIVKOVIC, KRISTOFFER RYPDAL}

\titlerunninghead{ON THE MAGNETOSPHERIC ORGANIZATION}

\authoraddr{T.  \v{Z}ivkovi\'{c},
Department of physics and Technology, University of Troms{\o},
9037 Troms{\o}, Norway
(tatjana.zivkovic@uit.no)}

\authoraddr{K. Rypdal, Department of physics and Technology, University of Troms{\o},
 9037 Troms{\o}, Norway}

\begin{document}

\title{Organization of the magnetosphere during  substorms}

\authors{T.  \v{Z}ivkovi\'{c}, \altaffilmark{1}
and K. Rypdal, \altaffilmark{1}}

\altaffiltext{1}{Department of Physics and Technology,
University of Troms{\o}, Norway.}

%
%


\begin{abstract}
The change in degree of  organization of the magnetosphere  during  substorms is investigated  by analyzing various geomagnetic indices, as well
as interplanetary magnetic field {\it z}-component and solar wind flow speed. We conclude that the magnetosphere self-organizes globally during substorms, but neither the magnetosphere nor the solar wind become more predictable in the course of a substorm. This conclusion is based on analysis of five hundred substorms in the period from 2000 to 2002. A minimal dynamic-stochastic model of the  driven magnetosphere that reproduces many statistical features of substorm indices  is discussed.
 \end{abstract}

\begin{article}

\section{Introduction}
The complexity of the magnetosphere has been extensively studied and one of its descriptions is based on the paradigm of self-organized criticality (SOC) coined  by  \cite{B87}. Systems that exhibit SOC activate many degrees of freedom, their
interactions are local, but due to the excitation of a scale-free hierarchy of avalanches when the system is slowly driven to a criticality  threshold, long-range interactions develop, and the dynamics on different spatial and temporal scales is essentially the same. It has been shown that the magnetospheric-ionospheric system exhibit  signatures  characteristic of SOC-dynamics \citep{VO91,U98}, and models for such description have been developed \citep{Valdivia,KlimasValdivia, C98, C99,Ur02,Koz03}. On the other hand, the magnetospheric-ionospheric system has also some signatures of intermittent turbulence \citep{Golov}, non-equilibrium phase-transitions \citep{S2001} as well as low-dimensional  chaos \citep{SVP93}.
It has been recently reported in \cite{TZ2011} that the magnetosphere, interplanetary magnetic field (IMF) {\it z}-component $B_{z}$ and solar wind flow speed $v$ become more predictable and more persistent during magnetic storms, while only the magnetosphere reduces the effective number of
degrees of freedom through  self-organization. In that study the magnetosphere was studied through analysis of the geomagnetic indices SYM-H and $D_{st}$, since it is known that these indices respond to the intensification of the ring current during magnetic storms \citep{W2006}.
In this article we analyze how different magnetospheric indices respond to the much more frequent and short-living events called substorms.

Different geomagnetic indices represent different parts of the magnetosphere and respond to different dynamics. Geomagnetic indices studied in this article, are downloaded from World Data Center, with 1-min resolution.
The most commonly used index for substorm studies is  the auroral electrojet index (AE) defined as the difference between the AU index, which measures the strength of the eastward electrojet in the auroral zone, and the AL index,  measuring the westward electrojet current, and is usually derived from 12 magnetometers  positioned below
the auroral oval \citep{DS66}.
We also use minute data for the IMF component $B_{z}$, as well as minute data for the solar wind bulk velocity $v$ along the
Sun-Earth axis. They are both retrieved from the OMNI satellite database and are given in GSE coordinate system. 

We have not analyzed other plasma parameters, since the plasma instruments can be influenced by solar X rays or energetic particle precipitation and are more unstable than the magnetometers.
We also analyze the polar cap magnetic activity (PC), as well as the AU and the AL index. During magnetically disturbed times the westward electrojet, whose proxy  is the AL index, increases abruptly due to currents from the magnetotail. On the other hand,  the eastward electrojet, whose proxy is the AU index, increases due to the partial ring current closure via the ionosphere in the evening sector \citep{Fe06}. Therefore, through the analysis of the AL and AU index
we can get  insight into the dynamics of different parts of the magnetosphere.
The PC index monitors geomagnetic activity over the polar caps caused by changes in the IMF and the solar wind.
This index is mostly influenced by field-aligned currents which flow at the poleward rim of the auroral oval, and is also sensitive to the
ionospheric Hall currents in the polar cap \citep{V91}, which are particularly dominant in the summer time. Since field-aligned currents are closely related to the auroral electrojets, the linear correlation of the PC and AE indices is of the order of 0.8-0.9
\citep{V91}. Here we use the northern polar cap index measured at the Danish geomagnetic observatory in Thule ($86.5^{\circ}$N).

Magnetic substorms are associated with  release and storage of  energy and momentum from the solar wind to the magnetosphere. They consist of three phases. In  the growth phase, typically lasting for about one hour,  loading of the magnetic flux and energy into the magnetotail takes place. It is succeeded by the expansion phase (substorm onset), lasting 30-60 minutes, when hot plasma ``unloads" earthward,  leading to sudden brightenings of the polar aurora, and plasmoids are ejected away from Earth in the far tail. In this phase dissipation is dominant and  formation of a storm ring current can take place. In the ionosphere substorm onset is characterized by a rise of the westward electrojet current, which forms the substorm current surge with field aligned currents. The  
recovery phase returns the magnetosphere to its quiet state. The duration of this this phase is 1-2 hours. 

Data about the times of substorm onsets are found in  \cite{FM04}, whose database is from the period between 2000 and 2002. These substorms were detected by the
FUV instrument on the IMAGE spacecraft. Observations covered the peak of the last solar cycle. According to  \cite{FM04} a substorm onset is only accepted as a separate event if at least 30 minutes have passed after the previous event.
It is also required that  local brightening of aurora occurs and that the aurora expands to the poleward boundary of the auroral oval and spreads azimuthally
in local time for at least 20 minutes. The latter criterion excludes pseudo-breakups which do not develop into full substorms.

The remainder of the paper is organized as follows: section 2 gives a brief overview of the methods used in the analysis of our data.  Section 3 shows the results from application of these methods to the substorm data and section 4
is reserved for discussion. 
\section{Methods}
\subsection{Recurrence-plot analysis}
Recurrence-plot analysis was developed by \cite{EK} and is very useful in studies of short and non-stationary time series. A comprehensive review of the method and its applications can be found in \cite{RQA}. Substorm durations are at most a few hours and all indices show non-stationary behavior during the events. Recurrence-plot analysis is very suitable for handling such short non-stationary time series. The method is useful for low-dimensional  deterministic dynamical systems as well as for high-dimensional systems and for stochastic signals. It can also provide useful information about non-autonomous systems. An impression of  the versatility of this technique can be obtained from the special issue edited by \cite{Marwan2008}. For low-dimensional systems
\begin{equation}
 \frac{d{\bf z}}{dt}={\bf f}[{\bf z}(t),t],  \label{eq1}
  \end{equation}
recurrence plots are based on the recurrences of a trajectory ${\bf z}(t)$ on the $d$-dimensional attractor in a $p>d$-dimensional phase space. If the system is autonomous, i.e. no explicit time dependence of  the phase-space flow ${\bf f}[{\bf z}]$, and if
 the attractor of trajectory has dimension $d$, Takens'  time-delay method  \citep{T81} can be used to construct an $m>2d$-dimensional embedding space on which the attractor can be mapped continuously and one-to-one. The embedding space is constructed from the time series  $x_i=g({\bf z}[i\, \delta t])$, where $g({\bf z})$ is the measurement function. Here $t=i\, \delta t$, and $\delta t$ is the sampling time of the time series. 
The mapping is given by
\begin{equation}
{\bf x}_{i}=(x_{i},x_{i+\tau},\dots,x_{i+(m-1)\tau}), \label{eq0}
\end{equation}
where $\tau$ is the time delay. 
There are practical constraints on useful choices of the time delay $\tau$. If $\tau$ is much smaller than the autocorrelation time the  image of the attractor in the embedding space becomes  essentially one-dimensional. If $\tau$ is much larger than the autocorrelation time, noise may destroy the deterministic connection between the components of  ${\bf x}(t)$, such that our assumption that ${\bf z}(t)$ determines ${\bf x}(t)$ will fail in practice. A common choice of $\tau$ has been the first minimum of the autocorrelation function, but it has been shown that better results are achieved by selecting the
 time delay  as the first minimum in the average mutual information function \citep{DA}, which we also use in this article.
 
 The recurrence-plot analysis deals with the trajectories in the embedding space. If the original time series $x(t)$ has $N$ elements and a time delay $\tau$, we have a time series of $N-(m-1)\tau$ vectors ${\bf x}(t)$ for $t=1,\ldots,N-(m-1)\tau$. This time series constitutes the trajectory in the reconstructed embedding space.

The next step is to construct a $N-(m-1)\tau\times N-(m-1)\tau$ matrix $R_{i,j}$ consisting of elements 0 and 1. The matrix element $(i,j)$ is 1 if  the distance is $\Vert {\bf x}_i-{\bf x}_j\Vert \leq \epsilon $ in the reconstructed  space, and otherwise it is 0. 
The recurrence plot is simply a plot where the points $(i,j)$ for which the corresponding matrix element is 1 is marked by a dot. For a  deterministic system the radius $\epsilon$ is typically  chosen as small fraction  of the diameter of the reconstructed attractor, but varies for different sets of data. In our analysis we have used $10\% $ of the total extent of the set spanned by the trajectory analyzed in the embedding space. This is a rule of the thumb generally accepted in the recurrence-plot literature \citep{RQA}. The  results are not very sensitive to this choice, but issues arise if $\epsilon$ is too large (to crude resolution and practically no information)  or too small (poor statistics).

Dynamical systems with a large number of independent or weakly dependent  degrees of freedom can only be described either by large-scale numerical simulation or by stochastic methods. For such systems the phase-space attractor is also high-dimensional and cannot be mapped one-to-one onto a low-dimensional  time-delay  embedding space. Nevertheless, the evolution of the ``projection" of the phase-space vector onto the embedding space usually provides valuable information if the measurement function is carefully chosen to be sensitive to  variation of those degrees of freedom that are in focus of our interest (for instance we should use the  AE index rather than the $D_{st}$ index as measurement function if substorm  activity is studied). Some of this information can be discerned  from the recurrence plots, even if the recurring states are not recurrences of the full phase-space vector, but only of the projections. This is true also if the full set of differential equations is non-autonomous. The time-dependent forcing invalidates the embedding theorem, but we are not assuming a perfect embedding anyway. We just have to be aware that dynamical features we observe in the recurrence plot may well reflect the dynamics of the forcing and not only the internal dynamics of the unforced system.

Of particular interest is analysis based on the diagonal line structures of the recurrence plots. If the projected trajectory  has a tendency to repeat its path when it recurs to the same region in embedding space, recurrences tend to be localized along unbroken diagonal segments. The longer these segments are, the more predictable is the path. We define the average inverse diagonal line segment length $\Gamma$:
\begin{equation}
\Gamma \equiv \langle l^{-1} \rangle=\sum_{l} l^{-1}P(l)/\sum_{l} P(l), \label{eq4}
\end{equation}
where $P(l)$ is a histogram over diagonal lengths:
\begin{equation}
P(l)=\sum_{i,j=1}^{N} (1-R_{i-1, j-1} ) (1-R_{i+l, j+l}) \prod_{k=0}^{l-1} R_{i+k, j+k}.  \label{eq5}
\end{equation}
It was shown heuristically in \cite{TZ2011} that $\Gamma$ can be used as a proxy for the largest Lyapunov exponent in deterministic systems and as a measure of persistence in  stochastic systems. For a deterministic system we demonstrated the connection with the largest Lyapunov  exponent $\lambda$ by computing $\lambda$ and $\Gamma$ along a period-doubling route to chaos for the Lorenz system. By computing $\Gamma$ and the self-similarity exponent $h$ for numerically generated fractional Brownian motions (fBm) with $h$ in the range $0<h<1$ we established the relation 
$\Gamma=0.72-0.57 h$ in the range of persistent
Brownian motions ($0.5<h<1$). An fBm is a Gaussian stochastic process $\{x(t)\}$ which satisfies the self-similarity condition $x(t)\stackrel{d}{=}\lambda^{-h}x(\lambda t)$ for all  $\lambda$. Here  $\stackrel{d}{=}$ denotes identity in distribution. For a given time-series $h$ can be estimated from the variogram \citep{TZ2011}. For a chaotic, deterministic system the inverse of the largest Lyapunov exponent is a measure of predictability. For a persistent stochastic process $h$ is  also such a measure, and $\Gamma$ decreases with increasing $h$. Thus we suggest use the inverse of $\Gamma$ as a measure of predictability. Since signals of interest  are a mixture of deterministic and stochastic components, and many stochastic processes are neither   self-similar nor Gaussian, we do not consider $\Gamma$ as an absolute measure of predictability, but when a decrease of $\Gamma $ takes place in a time series under certain conditions, e.g. during a magnetic storm, we interpret this as increase in predictability of the dynamics under these conditions.

We compute $\Gamma$  for embedding dimension $m=1$. An obvious advantage of this choice is that we don't have to worry about the choice of time delay $\tau$. A higher value of  $m$ does not make much more sense since  the  dynamics of the magnetosphere and the solar wind  is strongly influenced by a high-dimensional/stochastic component and cannot be unfolded in any higher-dimensional embedding space. 
Also, if the embedding dimension is inappropriately high, spurious long diagonal lines will appear in the recurrence plot.
Moreover, the way  $\Gamma$ depends on other predictability measures like Lyapunov exponent or self-similarity exponent is rather insensitive to the choice of  $m$, hence $\Gamma$ can be used to detect changes in predictability irrespective of the choice of embedding dimension.
\subsection{A test for determinism}
We shall adopt a  terminology where a physical system is deterministic if it can be described as a low-dimensional dynamical system, i.e. by the autonomous version of equation (\ref{eq1}). A test of determinism   was developed by \cite{KG92,KG93} which takes advantage of the fact that the trajectory through a given position in phase space is completely determined by this position, and if a trajectory recurs to the vicinity of this point the tangents of the trajectory in these two points are approximately parallel. This is in contrast to a stochastic  system, where the directions of the tangents are independent for the two points. Let us assume that  the phase space is divided into boxes. In our test, a box size is  chosen as the average distance a phase-space point moves in the $m$-dimensional embedding space during one time step.
The displacement of the trajectory inside a box, in {\it m}-dimensional phase space is given from the time-delay embedding reconstruction: 
\begin{eqnarray}
\Delta {\bf x}(t)&=&[ x(t+b)-x(t), x(t+\tau+b)-x(t+\tau),\ldots , \nonumber  \\ 
& &x(t+(m-1)\tau+b)-x(t+(m-1)\tau)] , \label{eq6}
\end{eqnarray}
 where $b$ is the characteristic time the trajectory spends inside a box. In our test $b=1$ time step.
The tangent for the $k$th pass through box $j$ is the unit vector ${\bf u}_{k,j}=\Delta {\bf x}_{k,j}(t)/\vert \Delta {\bf x}_{k,j}(t)\vert$.
The  averaged tangent in the box is
 \begin{equation}
{\bf V}_{j}=\frac{1}{n_{j}}\sum_{k=1}^{n_{j}}{\bf u}_{k,j}, \label{eq7}
\end{equation}
 where $n_j$ is the number of passes of the trajectory through box $j$. In the case of deterministic dynamics and finite box size, ${\bf V}_{j}$ will not depend very much on the  number of passes $n_j$, and $V_j$ will  converge to $1$. 
 In contrast, for the trajectory of a random process with independent increments, $V_j$ will decrease with $n_j$ as $V_j\sim n_j^{-1/2}$. 
Thus, to obtain better statistics for the description of average tangents we compute the average $V_j$ as a function of the number of passes through a box:
 \begin{equation}
 L_n\equiv\langle  V_{j}  \rangle_{n_j=n}, \label{eq8}
 \end{equation} 
 where this average is done over all boxes with same number $n$ of trajectory passes.

There are obvious similarities between this test for determinism and the one   for predictability developed in the previous  subsection. Both measure the degree of divergence of projected trajectories starting at almost the same position in the reduced embedding space. In those cases when the system can be described as a low-dimensional dynamical system (equation (\ref{eq1}) without the explicit time dependence in the flow field), we have $L_n=1$ for all $n$, provided the embedding dimension $m$ is sufficiently large to unfold the attractor. In this case $\Gamma$ will reflect the magnitude of the largest Lyapunov exponent, and the two tests clearly measure different properties. For a stochastic process $L_n$ is independent of $m$ \citep{TZ2011} and  independent of its persistence, which is easily demonstrated  by computing it for fBms with varying self-similarity exponents $h$. However, for the systems considered in this paper the signals contain both a  deterministic and a stochastic component. In this case the phase-space trajectory cannot be mapped one-to-one onto the embedding space for any choice of $m$, and the distinction between the two methods is less clear. One could envisage that an increase of the low-dimensional (deterministic) component relative to the stochastic one  could enhance both $L_n$ and $\Gamma^{-1}$. On the other hand an increase in predictability in either the deterministic or the stochastic component, without change in relative strength of the two components, could also influence both measures. Thus, in those cases where both measures either increase or decrease together, it is difficult to decide whether the cause is a change in determinism or in predictability. However, in situations when only one of the measures undergoes a change, or they change in opposite direction, the interpretation is unambiguous.

\cite{KG92,KG93} also suggest another test of determinism, which does not suffer from this ambiguity.
It  takes advantage of how the measure of determinism, the curves $L_n$, changes when we construct a new surrogate time  series for which the effect of nonlinearity in a low-dimensional system has  been corrupted. This surrogate signal has the same power spectral density (and hence same auto-covariance) as the original signal, but  the phases of the Fourier coefficients have been randomized. From application of this procedure to synthetic stochastic and low-dimensional signals we have gained support for the conjecture that randomization of phases do not change these curves for neither stochastic or high-dimensional, nor low-dimensional, linear systems. Only for low-dimensional, nonlinear systems will the surrogate data curves lie below those based on the original data. These features were demonstrated  for numerical solutions of the Lorenz system (low-dimensional, nonlinear and chaotic)  in \cite{TZ2011}. 
Here we demonstrate the same for $L_{n}$ versus $n$ for the Mackey-Glass (MG) equation \cite{MG77}:
\begin{equation}
\frac{dx}{dt}=-bx(t)+a\frac{x(t-\delta)}{1+x(t-\delta)^c} \label{eq2}
\end{equation}
Unlike the Lorenz system, the attractor of this system is in general high-dimensional. However, for some set of parameters, e.g.  $a=0.2, b=0.1, \delta=100, c=10$, the attractor of MG dynamics can be low-dimensional and $L_{n}$ derived from the  MG equation will fall more slowly with increasing $n$ than that for the randomized version, as shown in Figure 1.
The inset in the same figure shows the same results for a fractional Ornstein-Uhlenbeck  (fO-U) stochastic process. The stochastic equation used to generate such a process  was described in \citep{TZ2011}. The coefficients in this equation  used to generate Figure 1 are generated from fitting the variogram of the numerically generated process to the variogram of the AE index by means of least square regression (for definition of the variogram see  equation (12) in  \cite{TZ2011}). 
In both fO-U and MG equation, the embedding dimension used is $m=8$.
Realizations of a stochastic process like fO-U is indistinguishable from realizations of a measurement function of  a high-dimensional deterministic system for which the embedding dimension is too small to unfold the attractor. The $L_n$-curve for the fO-U process is unchanged after randomization of phases and shows that this test for determinism is negative for stochastic (and high-dimensional)  systems. We have verified  that this is the case also for strongly persistent (highly predictable) fO-U processes. This is completely reasonable on theoretical grounds; the long-range persistence in an fBm only depends on the predominance of low frequencies in the power spectrum, not on correlation between the phases of the Fourier coefficients.  Thus, this  test is not measuring predictability and is  a useful test for detecting changes in determinism in time  series.
\begin{figure}
\begin{center}
\includegraphics[width=8cm]{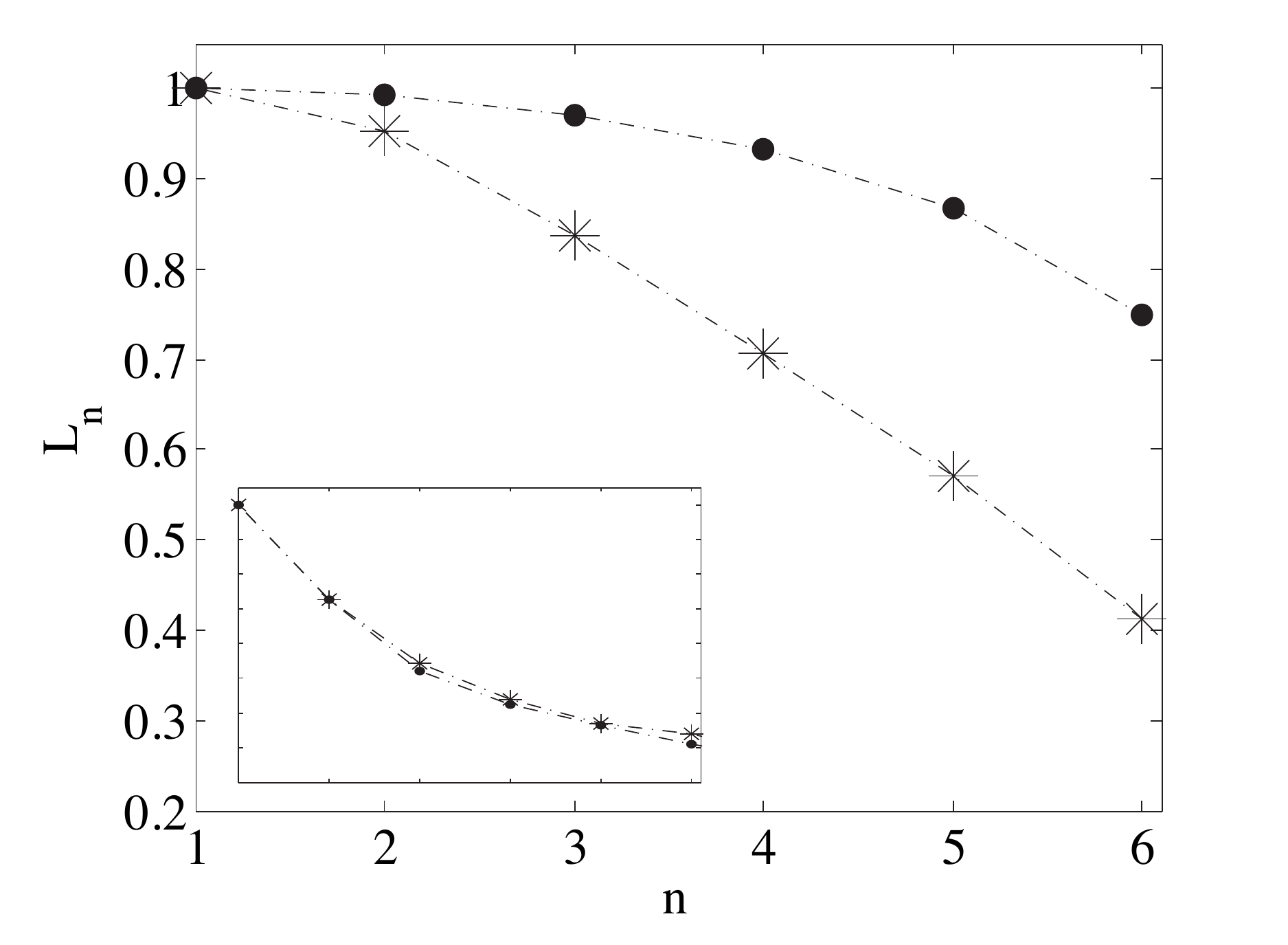}
\caption{Mean vector length $L_n $ vs. number of passages $n$: bullets are derived from numerical solutions to the MG equation, stars are derived from these solutions after randomization of phases of Fourier coefficients. Inset in the figure shows same characteristics for the fO-U process. The error bars are of the same order as the symbol size.}
\end{center}
\end{figure}
\begin{figure}
\begin{center}
\includegraphics[width=8cm]{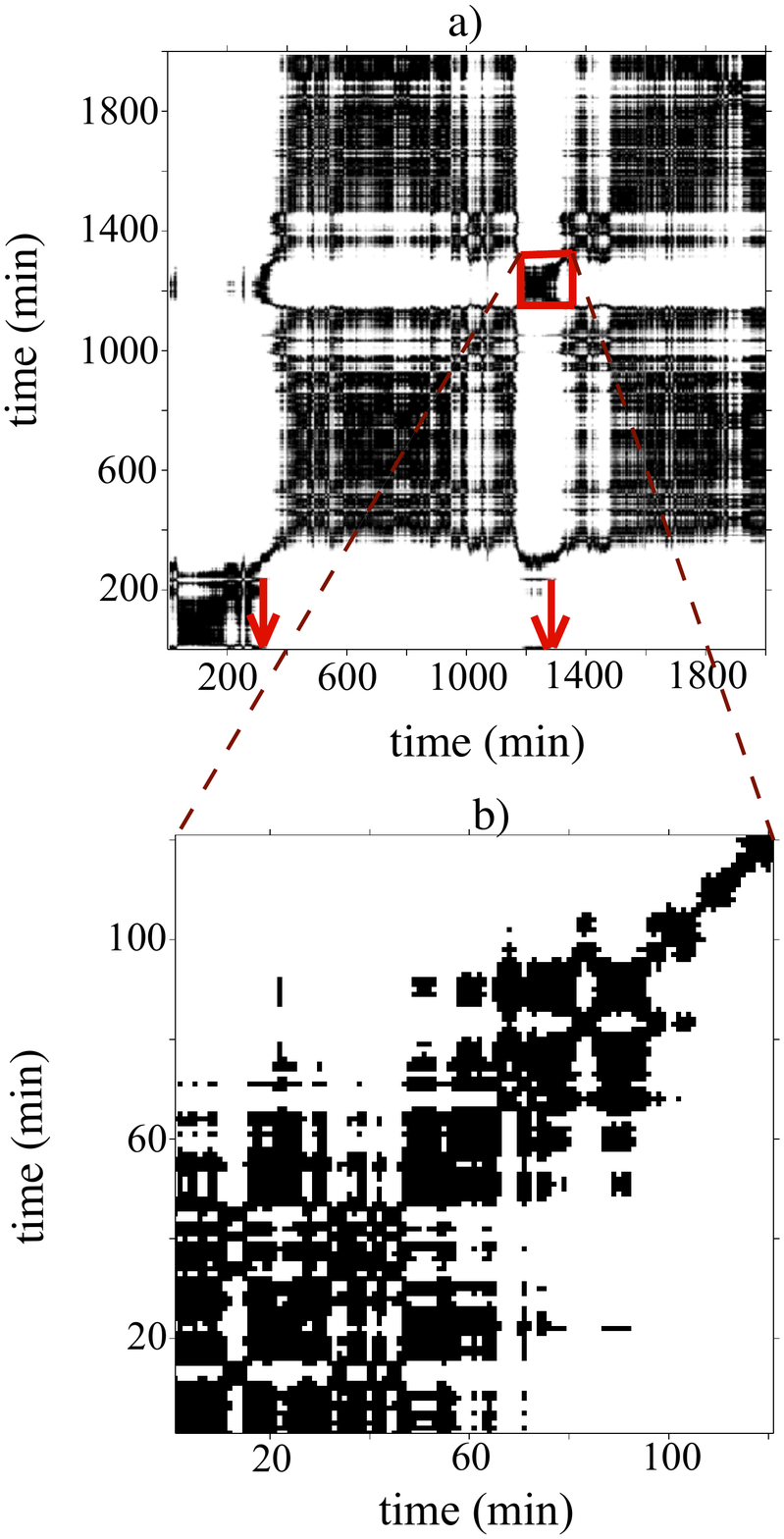}
\caption{Recurrence plots for tAE: a) for the entire day of September 8th, 2002, containing two distinct substorms whose onsets are marked by arrows. b) Blow-up of recurrence plot for second substorm (marked by square in panel (a)).}
\end{center}
\end{figure}
\section{Results}
In \cite{RR2010} it was shown that the fluctuation amplitude (or more precisely; the one-timestep increment) $\Delta y(t)$ of the AE index is on the average proportional to the instantaneous value $y(t)$ of the index. This gives rise to a special kind of intermittency associated with multiplicative noises, and leads to a  non-stationary  time series of increments. However, the time series $\Delta y(t)/y(t)$ is  stationary, implying that the stochastic process  $x(t)=\log y(t)$ has stationary increments. Thus, a signal  with stationary increments, which still can exhibit a multifractal intermittency,  can  be constructed by considering the logarithm of the AE index. 
We use transformed versions of geomagnetic indices: tAE$=\log(AE)$, tAL$=\log(0.1928\ \vert \mbox{AL}\vert+10.50)$, tAU=$\log(0.1167\ \vert \mbox{AU}\vert+13.50)$, tPC$=\log(0.0503\ \vert \mbox{PC}\vert+0.1978)$, while $B_{z}$ and $v$ have  increments which are not strongly dependent on their magnitude, and do not need transformation to obtain stationary increments.

It is well known that the AE index exhibits scale-free characteristics which often is associated with stochastic dynamics since its power spectral density has two distinct  power-law regimes \citep{T90}. However, it has also  been  shown that apart from colored noise which dominates the dynamics of the AE index on time scales up to 100 minutes, there are also signatures of  low-dimensional, chaotic characteristics as well \citep{AP2001}. These properties will be discerned from the analysis of determinism of geomagnetic indices in section 3.2.

\subsection{Predictability analysis}
\begin{figure} 
\includegraphics[width=7cm]{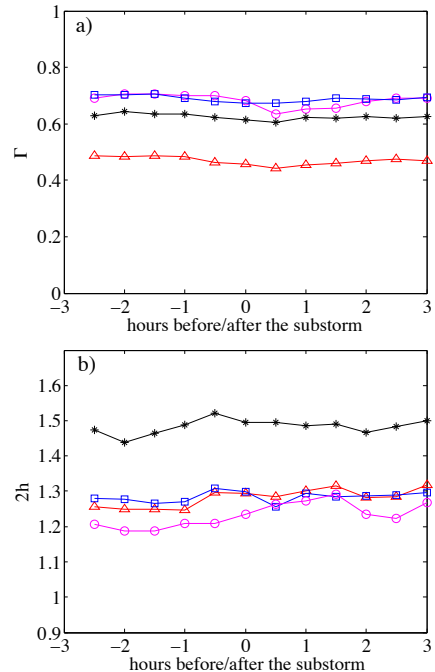} 
\caption{a) Ensemble mean of $\Gamma$ averaged over 500  substorms.  b) Ensemble mean of  $2h$. Triangles represent tAE, squares  tAU, stars  tPC, and circles  tAL. The standard errors for all points are in the range $0.02<\sigma< 0.04$. } 
\end{figure} 
First, we discuss two magnetic substorms that occurred on September 8th, 2002. The first substorm onset was registered at 5:01 UT and the other  at 21:26 UT.
The second substorm onset started after 13 hours of northward IMF $B_{z}$. 
In Figure 2a  the transition between dark and white bands in the recurrence plot indicates changes in the dynamics in the tAE. The first transition marked by an arrow occurs around the first substorm onset. The next  appears at the onset of the second substorm. For instance, the broad white band in Figure 2a corresponding to $i>300, j< 300$ indicates that a region in embedding space visited before the substorm onset is never visited after onset. Figure 2b contains a blow-up of the recurrence plot  for the second substorm. The image pales away from the main diagonal, which indicates that the process is non-stationary due to the intensification and the movement of ionospheric currents during the substorm. Since the plot pattern changes  during the  substorm we might
expect that $\Gamma$, which is defined from the diagonal lines of the plot, will change as well. 
\begin{figure} 
\includegraphics[width=7cm]{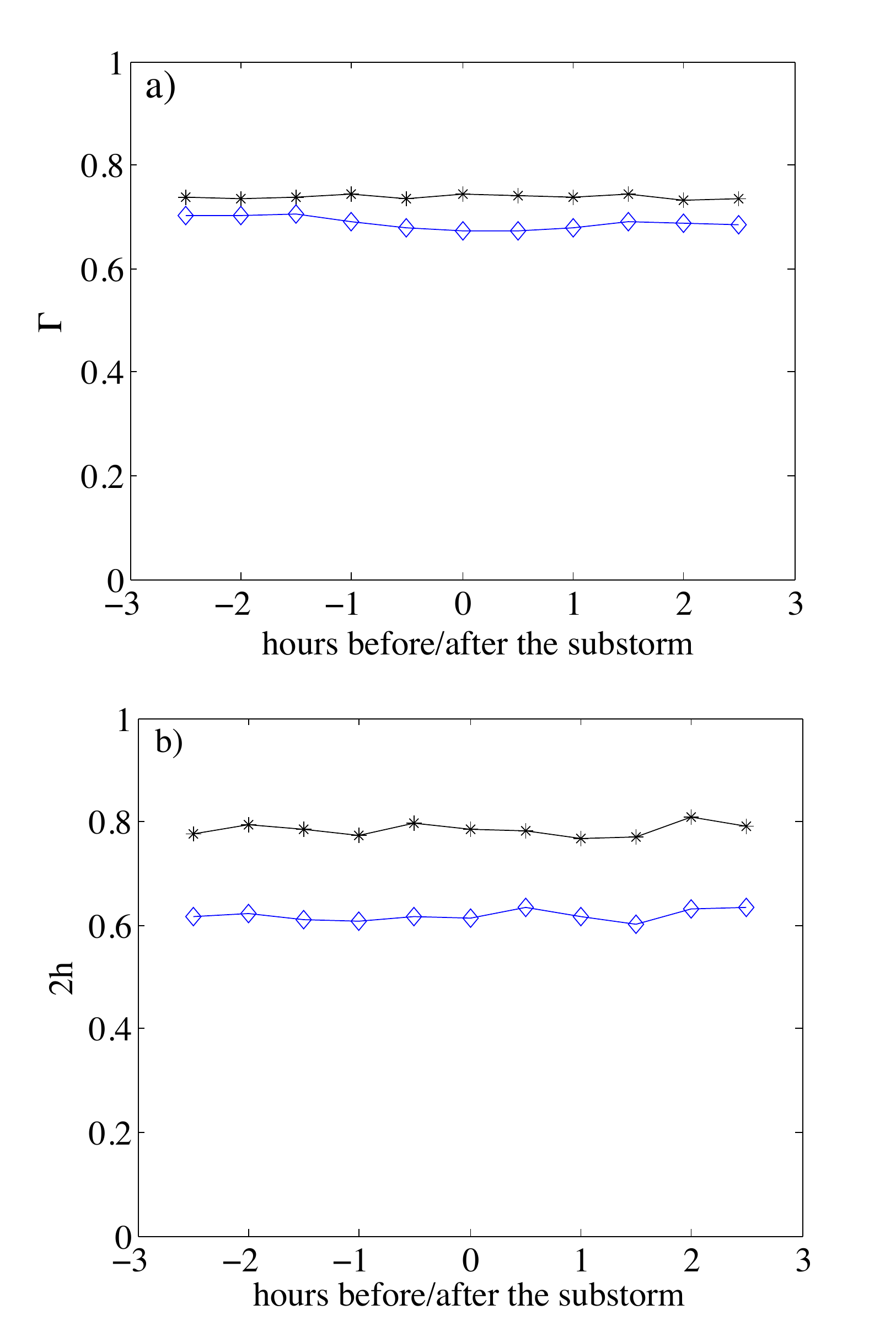} 
\caption{a) Ensemble mean for $\Gamma$. b) Ensemble mean  for $2h$. Stars represent $B_{z}$ and diamonds  $v$. Standard error for $\Gamma$ is of the order $\sigma \approx 0.02$, while for $2h$ it is $\sigma \approx 0.04$.} 
\end{figure} 
\begin{figure}
\begin{center}
\includegraphics[width=8cm]{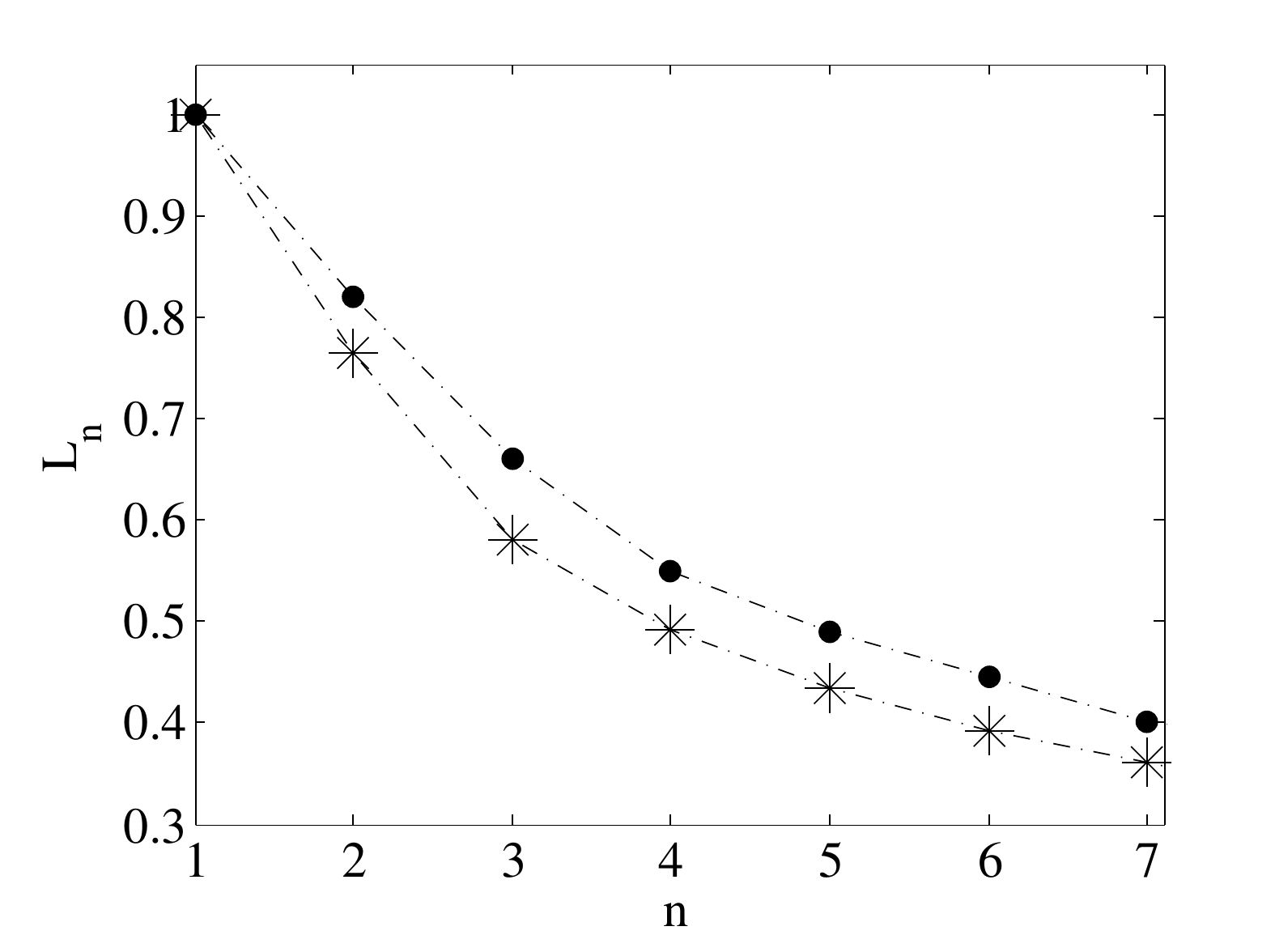}
\caption{Mean vector length $L_{n}$ vs. number of passages $n$: bullets are derived from the entire tAE time series, stars  from this time series after randomization of phases of the Fourier coefficiens. Standard errors are of the same order as the symbol size.}
\end{center}
\end{figure}
We obtain $\Gamma$ for tAE, tAU, tPC, tAL, $B_{z}$, and $v$ for 500 substorms whose onsets were taken from the database of \cite{FM04}.  The variation of $\Gamma$ over a 6 hour time interval, three hours before and three hours after the substorm onset, is computed from recurrence plots derived from thirty-minute windows, such that 12 $\Gamma$-values  are obtained for each substorm.
 The time evolution of $\Gamma$ is computed for all 500 substorms and averaged. The 
results are presented in Figure 3a. Apparently,  there is no significant variation of the ensemble average of $\Gamma$ over the duration of a substorm for any of the geomagnetic indices.
\begin{figure}
\begin{center}
\includegraphics[width=8cm]{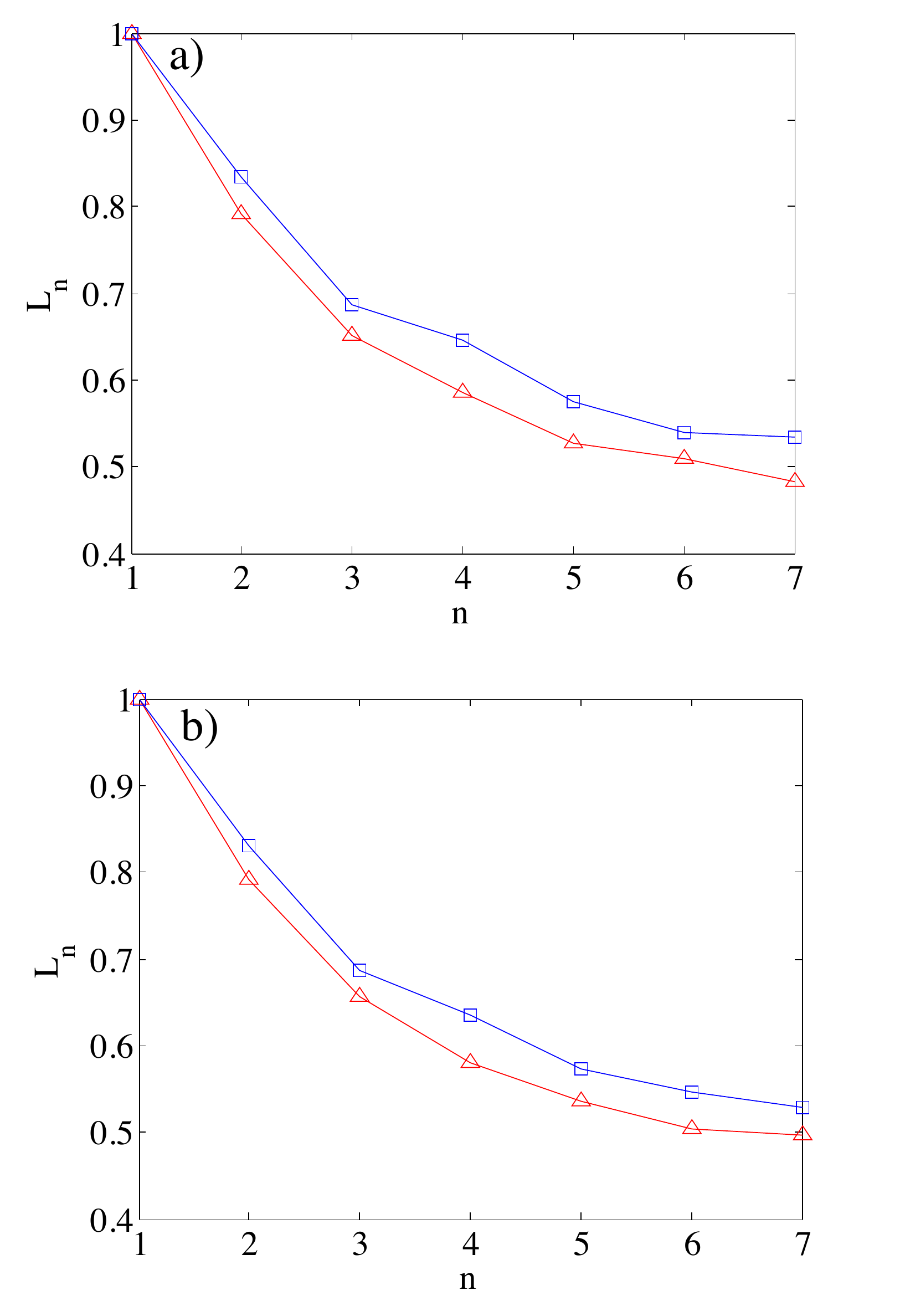}
\caption{$L_{n}$ vs. $n$, a) squares are derived from tAE during substorms, triangles  from the entire  tAE time series, b) squares  from tAU during substorms, triangles from entire  tAU time series.  Standard errors are of the same order as the symbol size.}
\end{center}
\end{figure}

\begin{figure}
\begin{center}
\includegraphics[width=8cm]{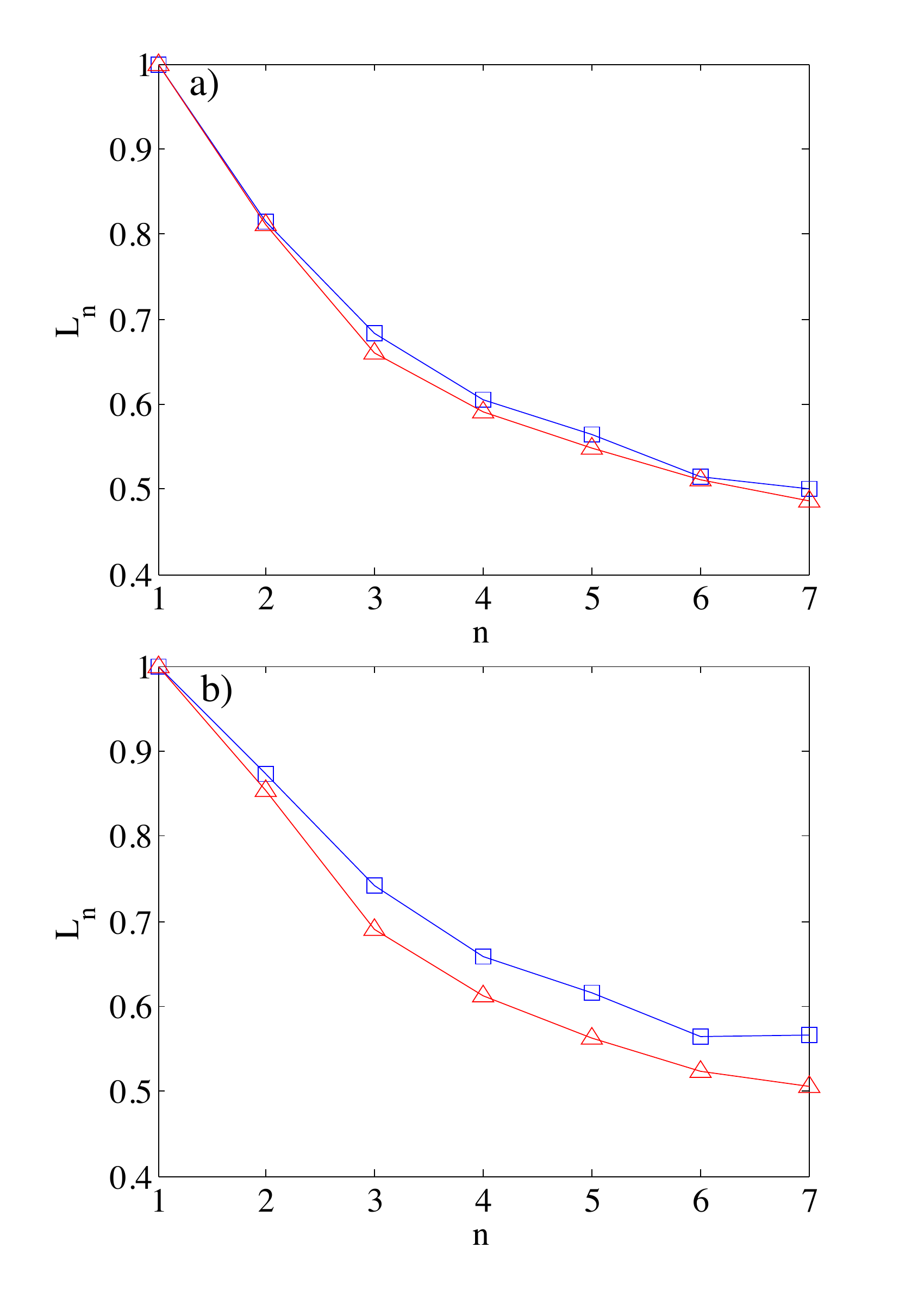}
\caption{$L_{n}$ vs. $n$, a) squares are derived from tAL during substorms, triangles  from the entire  tAL time series, b) squares  from tPC during substorms, triangles from entire  tPC time series. Standard errors are of the same order as the symbol size.}
\end{center}
\end{figure}
On time scales less than 100 minutes all quantities analyzed can be modeled as non-stationary multifractal motions whose persistence can be characterized by an exponent $h$ \citep{RR2010,RR2}.  Increasing $h$ implies higher persistence and predictability, corresponding to a reduction of  $\Gamma$. We compute $h$ from the variogram (equation (12) in  \cite{TZ2011}).
Figure 3b shows no variation in ensemble mean of $2h$ for any of the indices, confirming the result obtained for $\Gamma$.  

The same analysis is done for $B_{z}$ and $v$. Mean values for $\Gamma$ and $2h$ are plotted in Figure 4, showing no significant change of predictabilty during substorms for the solar wind observables. 
The example of two substorms from September, 8th, 2002 shows a reduction of $\Gamma$ in tAE around substorm onset (not shown here), but this day seems to be an exception rather than the rule.

 \begin{figure}
\begin{center}
\includegraphics[width=8cm]{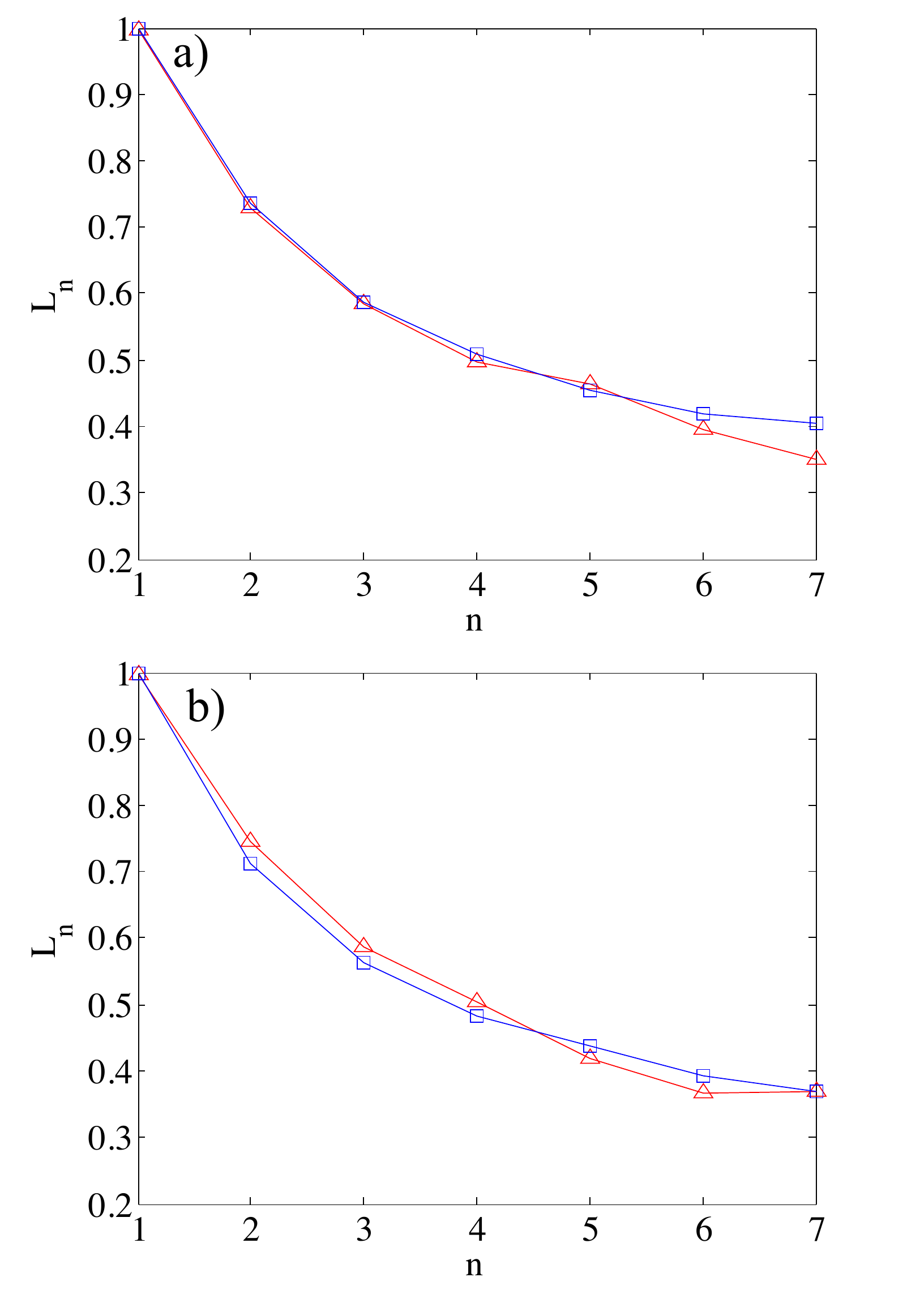}
\caption{$L_{n}$ vs. $n$, a) squares are derived from $B_{z}$ during substorms, triangles from entire $B_{z}$ time series, b) squares are derived from $v$ during substorms, triangles from entire $v$ time series. Standard errors are of the same order as the symbol size.}
\end{center}
\end{figure}

\begin{figure}
\begin{center}
\includegraphics[width=8cm]{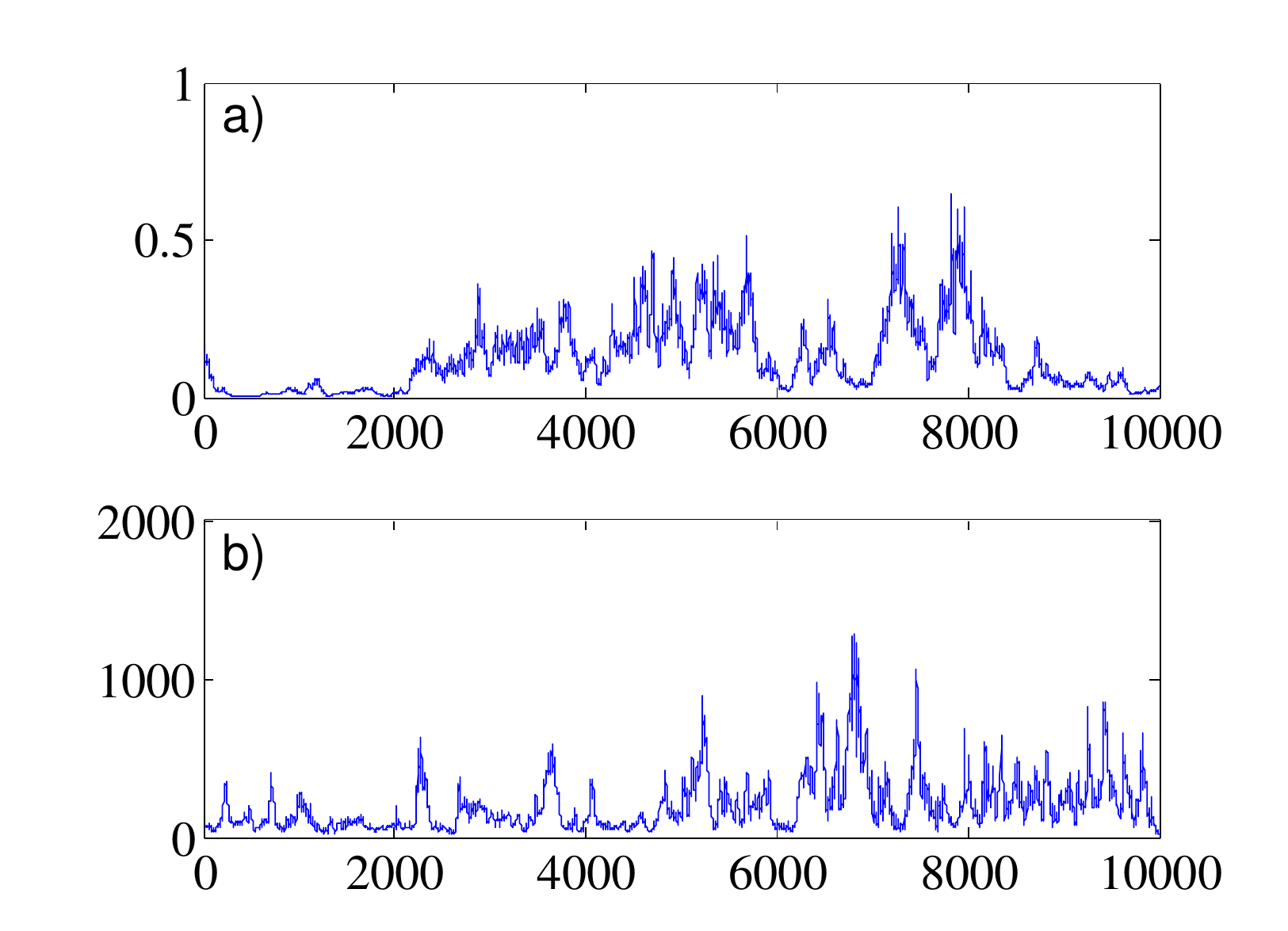}
\caption{a) Numerical solutions of equation (\ref{on-offeq}), b) AE index.}
\end{center}
\end{figure}
 \begin{figure}
\begin{center}
\includegraphics[width=8cm]{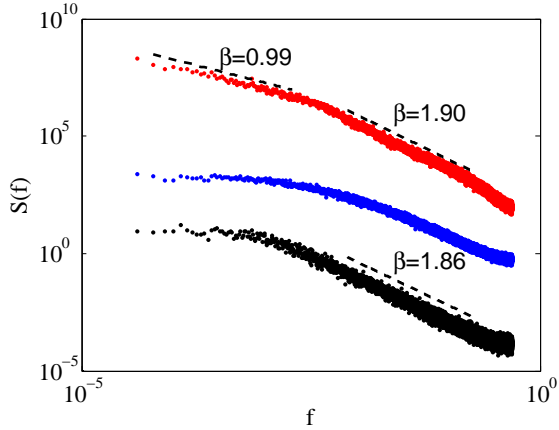}
\caption{Power spectral denities. Upper curve: from AE index.  Middle: from solutions of equation (\ref{on-offeq}) with $w$ a fractional Gaussian noise with $H=0.45$. Lower: from  fO-U process with $w$ the same as above.}
\end{center}
\end{figure}
 \begin{figure}
\begin{center}
\includegraphics[width=8cm]{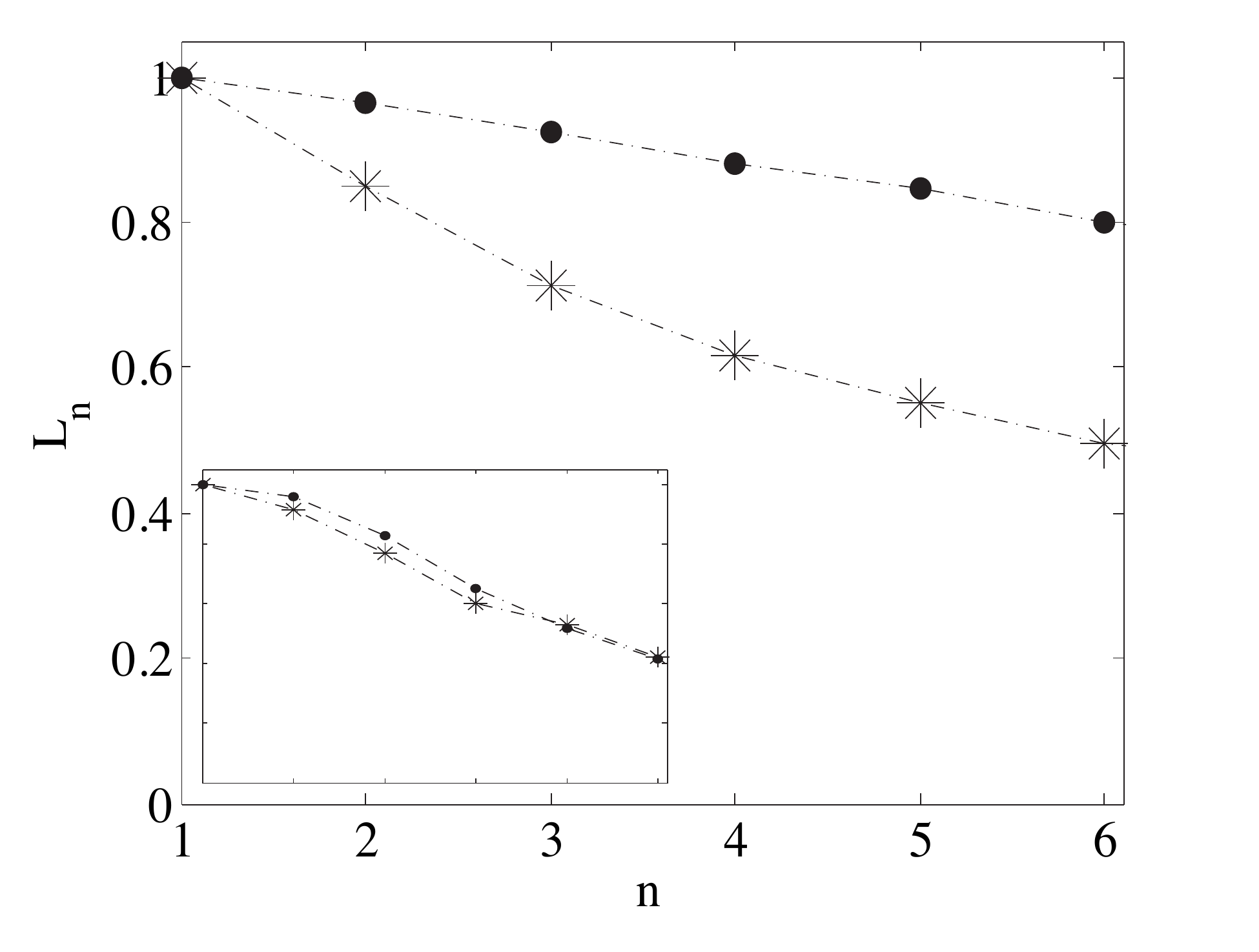}
\caption{Mean vector length $L_n $ vs. number of passages $n$: bullets are derived from low-pass filtered numerical solutions to  equation (\ref{on-offeq}), stars are derived from these solutions after randomization of phases of Fourier coefficients. Inset in the figure shows same characteristics for the filtered fO-U process. Standard errors are of the same order as the symbol size.}
\end{center}
\end{figure}
 \begin{figure}
\begin{center}
\includegraphics[width=8cm]{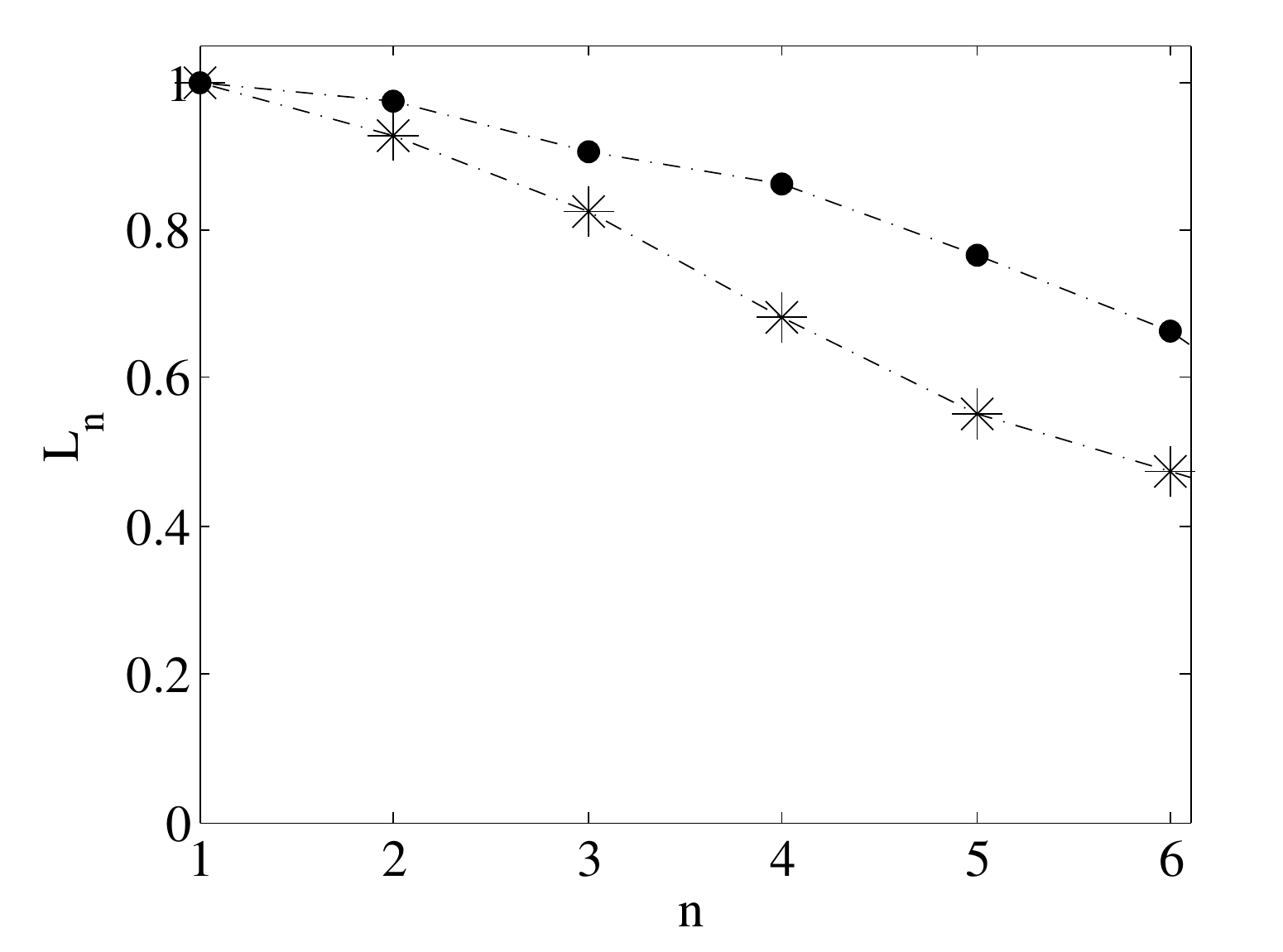}
\caption{Mean vector length $L_{n}$ vs. number of passages $n$: bullets are derived from the entire low-pass filtered tAE time series, stars  from this time series after randomization of phases of the Fourier coefficients. Standard errors are of the same order as the symbol size.}
\end{center}
\end{figure}

\subsection{Determinism analysis}
The signals we study are dominated by a stochastic component and their $L_{n}$ decreases when number of passes $n$ is increased. 
However, the existence of a low-dimensional component in e.g. the tAE can be demonstrated by computing $L_{n}$ and then do the same for the surrogate signals obtained by randomizing phases of Fourier coefficients.
In order to compute mean $L_{n}$, we use AE index data from the year 2000; each $L_{n}$ is computed over a record of 7200 points, such that the mean $L_{n}$ is computed over $N=70$ realizations. We have also computed mean $L_{n}$ for several
other lengths of records, but its value has not changed significantly. The embedding dimension is $m=6$, and $\tau=20$ minutes.
The sample values of $L_n$ are normally distributed, and hence the standard error of the ensemble mean is $\sigma/\sqrt{N}$.
This error is for all $L_n$-curves shown in this paper of the same order as the symbol size, and is therefore not shown explicitly in the figures.
As observed in Figure 5, $L_{n}$  for tAE is higher  than for the randomized version, indicating  existence of a  low-dimensional and nonlinear component in the signal. The same is shown for tAL, tAU, and tPC indices.
Notice that in case of fO-U in the inset of Figure 1, $L_n$ for the randomized version could not be distinguished from the $L_{n}$ of the fO-U itself.

Also, the same test was done for the geomagnetic index SYM-H in \cite{TZ2011}, and even though analysis of  this index 
yields low-dimensionality during magnetic storms, $L_{n}$ is  indistinguishable from its randomized version when averaged over a year. $L_{n}$ computed for $B_{z}$ and $v$ is also indistinguishable from the $L_{n}$ of their randomized versions (not shown here). 

As we shall demonstrate below tAE, tAU and tPC all exhibit a discernible low-dimensional component during  substorms. Since  substorms occur very often, sometimes several per day, this low-dimensionality is discernible also when averaged over the year (as is shown in Figure 5).  We use the substorm database and compute $L_{n}$ in the time interval 1 hour before and 1.5 hours after  the substorm onset. This way we generally include all phases of the substorm.

It has been shown in \cite{TZ2011} that the increase in the embedding dimension gives increased determinism for the systems with low-dimensional dynamical component, while no change in determinism can be seen for the fO-U process. Therefore, the only constrain on the embedding dimension for our data should be the length of the studied time series.
Ideally $\tau$ should be chosen to correspond to the first minimum of the average mutual information (which is $\tau \approx 20$ in most of our data). However, in our data we are limited by the length of the time records, which is the typical length of a substorm. We use records of length 2.5 hours (record length $L=150$ data points), chose $\tau=10$ minutes, and $m=6$ (when $\tau=20$, results are the same, but due to better statistics, we use $\tau=10$).
This leaves a trajectory record in embedding space  of length $N-(m-1)\tau =100$ (from equation (2)), from which we compute a sample value of $L_n$ for a given substorm. These values are computed over an ensemble of $N=500$ substorms; and the ensemble mean and standard error $\sigma$ are estimated. 

This analysis is made for tAE, tAL, tAU, and tPC, and  compared to $L_{n}$ computed from all the data from year 2000. Figures 6 and 7 illustrate that all geomagnetic indices (except tAL) exhibit elevated  determinism during substorm times, i.e. low-dimensionality increases during substorms. Future studies should investigate why determinism in tAL index does not change during magnetospheric substorms. Notice that if we could obtain equally good statistics with higher embedding dimension, it would be very likely that the determinism during substorms would be even more elevated than it is for $m=6$.

Another way to  demonstrate the existence of a low-dimensional  component in these indices is to compare the determinism computed from the tAE during substorms  with signals generated numerically from the fO-U equation, whose coefficients are fitted
to the tAE under substorm through least square regression procedure.
For each of a large number of realizations of the fO-U process, we compute the ensemble mean of  $L_{4}$ as a measure of low-dimensionality. 
The choice of $L_{4}$ from the $L_{n}$ vs. $n$ curve is a compromise between clear separation between low-dimensional and stochastic dynamics and small error bars (which increase with increasing $n$).
The ensemble mean for fO-U process is $L_{4}\approx 0.40$ , while for deterministic process  $L_{4}\sim 1$.  In contrast, Figure 6a shows that $L_{4}\approx 0.65$ for tAE under substorm conditions, which gives a clear indication that  the AE index is neither fully deterministic nor stochastic.

 In Figure 8 we plot $L_{n}$ for $B_{z}$ and $v$ and observe no increase in determinsm  during substorms.  This indicates that the increased low-dimensional component in the geomagnetic indices during substorms is not imposed on the magnetosphere by the growth of such a component in the solar wind. In other words: the organization of the magnetosphere during substorms is a  {\em self}-organization. 
 
 \section{Discussion}
 
We have applied recurrence-plot analysis and a test of determinism to geomagnetic indices AE, AU, AL and PC, as well as IMF $B_{z}$ and solar flow speed $v$. Recurrence plots were applied by \cite{smorovi} to connect solar wind observables to the AL and  AU indices, concluding that the correlation between these indices results from magnetic storm signatures appearing in both time series. 
Also, correlation between the solar wind electric field $vB_{z}$ and the AE index was studied in \cite{CMI},  invoking mutual information. Here it was found that  correlation is present intermittently on the timescales of a few hours, suggesting that substorms are information carriers between the solar wind and the magnetosphere.
In our study, recurrence analysis is used to measure the inverse average diagonal line length $\Gamma$ which is heuristically shown to be a useful measure of predictability of the dynamics. As an alternative predictability measure the self-similarity exponent $h$ was measured during  the course of substorms.  No systematic variation of $h$ or $\Gamma$ during substorms  has been detected in an ensemble of five hundred substorms. This is true both for geomagnetic indices and the solar wind observables. Thus we conclude that geomagnetic indices which represent dynamics  in the magnetotail, plasma sheet boundary layer, partial ring current and polar cap convection, do not become more predictable during substorms. The same applies for observables representing the dynamical state of the solar wind driver.
From the analysis of a spatio-temporal dynamical model of the high latitude magnetic perturbation, it was shown in \cite{Valdivia99} that the nonlinear dynamical model for the evolution of the spatial structure produces better prediction than the linear one during
intense magnetospheric activity.
In this article, a simple test for determinism has shown as well that the AE index exhibits some nonlinear characteristics, but in addition, it has s weak low-dimensional component. These characteristics are elevated during substorms.
Other geomagnetic indices, except the AL index, also show same signatures during magnetic substorms, which could indicate that the magnetosphere self-organizes
and develops low-dimensional dynamics under substorm conditions. 
For the AE index similar results were obtained by \cite{AP2001}, who applied singular value decomposition (SVD) analysis to the AE index. They concluded, by comparing the AE index with solutions of the Lorenz system contaminated by a colored noise term, that the first SVD component in both cases is entirely due to colored noise, while the higher-order SVD components are due to the internal, low-dimensional and chaotic  dynamics. By computing cross-correlation between the first SVD component of the AE index and the index itself, they concluded that the influence of  the first component is about  40 percent. Further, the first SVD component is attributed to the solar wind and is characterized as linear and stochastic, while higher SVD components are attributed to  magnetospheric dynamics. In our test of determinism  the  data are not filtered to reduce the effect of the stochastic component on the  analysis, but the results still reveal that unlike IMF $B_{z}$ and solar wind flow speed $v$, the AE index contains components that makes it different from a high-dimensional or  stochastic system. 

Thus, it seems firmly established that the auroral electrojet proxies exhibit an additional low-dimensional component associated with self-organization of the magnetosphere, the most prominent example being the auroral substorm. It should be stressed, however, that also for AE activity the {\em major} component is stochastic.  \cite{CW2001} argue that the AE index contains  information transferred from the solar wind, since it has been shown in \cite{F} that power laws in burst lifetime distribution for the AL and AU indices and for the solar wind electric field $vB_{z}$ are similar
apart from a bump in the lifetime distribution which \cite{F} explain as a signature of the substorm current system. This  indicates that the stochastic component of the auroral electrojet activity to a great extent is a direct imprint of the solar wind turbulence. This conclusion is supported by the analysis in \cite{RR2}, where  $B_z$ and tAE are found to exhibit very similar multifractal spectrum on the time scale up to 100 minutes. 

Our analysis supports a picture where the magnetosphere under quiet conditions resides in a forced state where the organization of the magnetospheric dynamics reflects the organization of the solar wind driver, i.e. the stochastic properties of the global magnetospheric system and the driver are very similar. If substorms are trigged by solar wind features, this trigger is not an increase in organization or predictability of the solar wind dynamics. During substorms geomagnetic indices are influenced by a self-organization that involves the major current systems in the magnetosphere-ionosphere, and hence indicates that a global instability is exited. This picture is consistent with the scenario described in \cite{C99}, and conceptually simpler forerunners like \cite{Lewis}. The latter sets out to explain the unpredictability of substorm onset, noting that   external triggers, like reversal from northward to southward pointing $B_z$, are not always identifiable, and  series of substorm cycles can occur if southward IMF persists for prolonged times. Southward IMF  opens up for  loading of magnetic flux and energy to the magnetosphere, and in \cite{Lewis}, and further elaborated by \cite{S2001},   parameters representing this loading are modeled as external time-dependent control parameters, and the magnetospheric state-variable (e.g. a geomagnetic index) is modeled via a non-autonomous system on the form
\begin{equation}
\frac{dX}{dt}=F(X,a,b), \label{nonauton} 
\end{equation}
where $a(t)$, $b(t)$ are two time-dependent external control parameters. The underlying assumption is  that the magnetosphere resides in a forced equilibrium corresponding to a stable fixed point of this system, and hence these equilibria are located on the surface $F(X,a,b)=0$ in the three-dimensional $(X,a,b)$-space. Choosing a third-order polynomial form, e.g.  $F(X,a,b)=X^3+aX+b$, gives rise to a folded  surface that opens the possibility of a cusp catastrophe which is interpreted as the substorm onset (expansion phase). According to this non-autonomous model for evolution of stable, forced equilibria the substorm onset is not  really unpredictable since it depends on the control parameters $a$ and $b$. However, the catastrophe can occur along a curve in the $(a,b)$-plane which may be hard to identify from this kind of conceptual model, and hence practical prediction may be difficult.

 One unsatisfactory feature of non-autonomous models of this kind is that the control parameters are not directly related to the state of the solar wind driver, but rather a result of the solar-wind/magnetosphere interaction that is an integral part of the  dynamical system to be modeled. Thus, in this respect a more satisfactory approach is to model the magnetosphere-ionosphere as a dynamical system which is autonomous under constant forcing. Such models can be constructed with varying degree of sophistication, \citep{Baker,Klimas,Horton}, and may give rise to chaotic signals that reproduce many of the random characteristics of the magnetospheric time series. 
However, being deterministic and  low-dimensional they cannot reproduce the strong stochastic component in the observational signals. On the other hand, such models can be generalized to include stochastic forcing from small-scale internal dynamics and deterministic and stochastic forcing from variations of the  driver. Thus, one may  conceive complicated as well as simple conceptual dynamic-stochastic models that can capture the essential stochastic dynamics as it presents itself in the observables studied here. In \cite{RR2010,RR2} the AE index is modeled by a simple dynamic-stochastic equation that reproduces the general   statistical features. The deterministic dynamics in this equation was represented by a drift term (a nonlinear damping) which prevents the solution to drift off to infinity, but the focus in those studies was on the time scales less than a 100 minutes, where the effect of the  drift term is small. A version of this stochastic difference equation that exhibits on-off intermittency for certain choices of parameters is 
\begin{equation}
\delta X=M(x)\delta t+\sqrt{D}X\, w \label{on-offeq} 
\end{equation}
where $M(x)=aX-X^3$ with $a>0$  models the  drift term found  from the AE index in \cite{RR2}. Here $\delta t$ is the discrete time step (the sampling interval of the time series) and $\delta X(t)=X(t+\delta t)-X(t)$.  In \cite{RR2010} $w(t)$ is modeled as a particular  multifractal stochastic  noise process  with unit variance. It was shown by \cite{Aumaitre} that the on-off intermittency of this equation is sensitive to the nature of the noise term. If $w(t)$ is approximated by a white Gaussian noise, equation (\ref{on-offeq}) in the limit $\delta \rightarrow 0$ reduces to the It\^{o} stochastic differential equation $$dX=(aX-X^3)\, dt+\sqrt{D}\, X dB(t),$$where $B(t)$ is the Wiener process (Brownian motion). It can be shown from the associated Fokker-Planck equation that the stationary probability density for $X$ is $P(X)=CX^{(2a/D)-1}e^{-(X^2/D)}$. The divergence of $P(X)$ as $X\rightarrow 0$ for $ 2a/D<1$ is due to the on-off intermittency in this regime, which makes the solution reside in the vicinity of $X=0$ for a considerable portion of the time, while for $2a/D-1$ small, but positive,  the solution has  an intermittent character more similar to the behavior of the AE index. Such a solution, for $D=0.1$, is shown in Figure 9a,  with a sample of the AE index shown in panel (b). Here $w$ was chosen as a weakly anti-persistent fractional Gaussian noise with Hurst exponent $H=0.45$, which is the $H$-value derived from the power spectral density for  AE shown in  Figure 10. 
The relation between the spectral index $\beta$, which is the slope of the straight lines fitted to the spectra plotted in a log-log plot, and the Hurst exponent of the differentiated signal is $\beta=2H+1$.  On time scales $<100$ time steps (minutes)  AE index has $\beta \approx 1.90$, corresponding to $H\approx0.45$. The power spectral density for the model signal shows a less clear power-law regime on these time scales, which is mainly due to a crude model for the nonlinear drift term $M(X)$ in equation (\ref{on-offeq}). A more ``box-like" function would remedy this.  On time scales $>100$ minutes the spectrum for the AE index has a pink-noise character ($\beta\approx 1$), while the model time series has a more gradual transition towards white noise. Further study on refinements of equation (\ref{on-offeq}) is required to settle whether these spectral features are possible to reproduce within this class of one-dimensional stochastic equations.

An interesting question is whether modeling along these lines can produce bursts (substorms) which enhance the determinism compared to a completely random process. As discussed before, the test based on comparing the $L_n$-curve with the one obtained after randomization of phases is a test on the existence of a low-dimensional and nonlinear component in the dynamics.  Direct computation of $L_n$ from the model signal does not reveal a significant reduction of $L_n$ after randomization, in contrast to what was found from the MG system signal in Figure 1. One obvious reason for this is that the signal from equation (\ref{on-offeq}) contains a strong stochastic component which is not present in the signal from the MG system. Hence, in this case it is necessary to perform a mild low-pass filtering of the model signal before computation of $L_n$ \citep{KG93}. The  result after filtering is shown in Figure 11, indicating the presence of a low-dimensional nonlinear component. To make sure that this filtering does not introduce spurious low-dimensional nonlinearities in the signal, we perform the same test to a filtered fO-U process with similar spectral characterics as our model signal (the power spectral density for the fO-U signal is displayed in Figure 10). The filtered fO-U process shows very small change of $L_n$ after randomization of phases, as shown in the inset of Figure 11. Obviously, low-pass filtering also have an effect on this test when applied to physical signals with a stochastic component. In Figure 12 we show the effect of applying the same filter to tAE, which should be  compared to the result for the unfiltered signal  shown in Figure  5.

The  nonlinearity  producing  determinism in the model signal from equation (\ref{on-offeq}) is a combination of the nonlinear drift term and the multiplicative noise term $\sqrt{D}Xw$.  The multiplicative term can be eliminated by the transformation $Y=\log X$, but then It\^{o}'s formula \citep{Gardiner} yields a new drift term on the form $\tilde{M}(Y)=e^{-Y}M(e^Y)-D/2$. Note that if our original drift term is a linear damping $M(X)=-\nu X$ the transformed drift term  reduces to a negative constant $\tilde{M}(Y)=-(\nu+D/2)$. This yields $Y(t)\rightarrow -\infty$, and hence $X(t)\rightarrow 0$ as $t\rightarrow \infty$, and  demonstrates that the drift term must be nonlinear  to produce stationary time series from a  model with a multiplicative noise term like equation (\ref{on-offeq}).

\begin{acknowledgments}
The authors acknowledge extensive discussions with M. Rypdal. Recurrence plot and the longest diagonal line are computed by means of the Matlab package downloaded from \begin{verbatim} http://www.agnld.uni-potsdam.de/~marwan/toolbox/. \end{verbatim}
 The authors would like to thank the Kyoto World Data Center for AE, AL, AU and PC indices.
\end{acknowledgments}

\end{article}

\end{document}